\documentclass[12pt,tightenlines,eqsecnum,floats,aps,amsmath,amssymb,nofootinbib,prd,showpacs]{revtex4}

\usepackage{setspace}
\usepackage{amsmath,amssymb,amsfonts,amsthm}
\usepackage{graphicx}
\usepackage{enumerate} 
\usepackage{colordvi} 

\def\be{\begin{equation}}
\def\ee{\end{equation}}
\def\ba{\begin{eqnarray}}
\def\ea{\end{eqnarray}}

\def\p{\partial}
\def\H{{\cal H}}
\def\Hkg{\H_{\rm kin}^{\rm grav}}

\def\h{\hat }

\def\su{{\rm su}}
\def\SU{{\rm SU}}
\def\SO{{\rm SO}}
\def\A{{\mathcal{A}}}
\def\Ab{\bar{\A}}

\def\cm{\rm cm}

\def\a{\alpha}
\def\b{\beta}
\def\g{\gamma}
\def\ep{\epsilon}
\def\l{\ell}
\def\lp{{\ell}_{\rm Pl}}
\def\lo{{\ell}_o}
\def\R{\mathbb{R}}
\def\S{\mathbb{S}}

\def\q{{}^o\!q}
\def\e{{}^o\!e}
\def\w{{}^o\!\omega}
\def\mb{\bar{\lambda}}
\def\lo{\ell_o}
\def\rmin{\rho_{\mathrm{min}}}
\def\rmax{\rho_{\mathrm{max}}}

\def\b{$\bullet\,\, $}
\def\f{\frac}
\def\dd{\textrm{d}}

\def\WDW{WDW\,\,}

\newcommand{\ket}[1]{\ensuremath{|#1\rangle}}
\newcommand{\ip}[2]{{\langle#1\,|\,#2\rangle}}
\newcommand{\rcr}{\rho_{\mathrm{crit}}}
\newcommand{\fs}[2]{{\textstyle\frac{#1}{#2}}} 

\usepackage{enumerate}

\usepackage{colordvi}
\newcounter{mnotecount}[section]

\newcommand{\comment}[1]{}

\begin{document}

\preprint{\vbox{\baselineskip=12pt \rightline{IGPG-07/01-02}
\rightline{NSF-KITP-07-05 }}}

\title{An Introduction to Loop Quantum Gravity through Cosmology}

\author{Abhay Ashtekar${}^{1,2}$}
\email{ashtekar@gravity.psu.edu} \affiliation{${}^{1}$Institute
for Gravitational Physics and Geometry, Physics Department, Penn
State, University Park, PA 16802, U.S.A.\\
${}^2$Kavli Institute for Theoretical Physics, University of
California, Santa Barbara CA 93107, USA}

\begin{abstract}

This introductory review is addressed to beginning researchers.
Some of the distinguishing features of loop quantum gravity are
illustrated through loop quantum cosmology of FRW models. In
particular, these examples illustrate: i) how `emergent time' can
arise; ii) how the technical issue of solving the Hamiltonian
constraint and constructing the \emph{physical} sector of the
theory can be handled; iii) how questions central to the Planck
scale physics can be answered using such a framework; and, iv) how
quantum geometry effects can dramatically change physics near
singularities and yet naturally turn themselves off and reproduce
classical general relativity when space-time curvature is
significantly weaker than the Planck scale.

\end{abstract}


\maketitle

\section{Introduction}
\label{s1}

My lectures at the first St\"uckleberg symposium began with an
introduction to the challenges of quantum gravity, then discussed
key features of loop quantum gravity (LQG) and finally summarized
applications to cosmology. However, in view of the page limit, in
these proceedings I decided to present the material from a
different perspective, through the lens of loop quantum cosmology
(LQC). Thus, rather than starting with general considerations and
then descending to applications, here I will follow an opposite
approach, using applications to illustrate some of the key
problems of quantum gravity and distinguishing features of LQG.%
\footnote{A brief history of quantum gravity can be found in
\cite{njp} and a detailed discussion of conceptual problems in
\cite{asbook}. A brief mathematical introduction to loop quantum
gravity can be found in \cite{alency} and much more detailed
reviews in \cite{alrev,crbook,ttbook}.}
The advantage of this reverse strategy is that students and other
beginning researchers can appreciate the key issues, challenges
and advances in a rather simple setting, without having to first
grasp the technical intricacies of full LQG. The main drawback is
that this approach leaves out some of the most interesting
developments such as statistical mechanical calculations of black
hole entropy, discussions of the issue of information loss, and
spin foams, particularly the recent advances in calculating
graviton propagators in the non-perturbative setting of LQG
\cite{cretal}.

With these caveats in mind, let us begin with a list of some of
the long standing conceptual and technical issues of quantum
gravity on which loop quantum cosmology has a bearing.

In general relativity, gravity is encoded in the very geometry of
space-time. The most dramatic features of general relativity can
be traced back to this dual role of geometry: the expansion of the
universe, the big-bang, formation of black holes and emergence of
gravitational waves as ripples of space-time curvature. Already in
the classical theory, it took physicists several decades to truly
appreciate the dynamical nature of geometry and to get used to the
absence of a kinematic background geometry. In quantum gravity,
this paradigm shift leads to a new level of conceptual and
technical difficulties.%
\footnote{There is a significant body of literature on this issue;
see e.g., \cite{asbook} and references therein. These difficulties
are now being discussed also in the string theory literature in
the context of the AdS/CFT conjecture.}

\begin{itemize}

\item The absence of a background geometry implies that classical
dynamics is generated by constraint equations. In the quantum
theory, physical states are solutions to quantum constraints. All
of physics, including the dynamical content of the theory, has to
be extracted from these solutions. But there is no external time
to phrase questions about evolution. Therefore we are led to ask:
Can we extract, from the arguments of the wave function, one
variable which can serve as \emph{emergent time} with respect to
which the other arguments `evolve'? If not, how does one interpret
the framework? What are the physical (i.e., Dirac) observables? In
a pioneering work, DeWitt proposed that the determinant of the
3-metric can be used as an `internal' time \cite{bd}.
Consequently, in much of the literature on the Wheeler-DeWitt
(WDW) approach to quantum cosmology, the scale factor is assumed
to play the role of time, although often only implicitly. However,
in closed models the scale factor fails to be monotonic due to
classical recollapse and cannot serve as a global time variable
already in the classical theory. Are there better alternatives at
least in the simple setting of quantum cosmology? If not, can we
still make physical predictions?

\item Can one construct a framework that cures the short-distance
difficulties faced by the classical theory near singularities,
while maintaining an agreement with it at large scales? By their
very construction, perturbative and effective descriptions have no
problem with the second requirement. However, physically their
implications can not be trusted at the Planck scale and
mathematically they generally fail to provide a deterministic
evolution across the putative singularity. In LQG the situation is
just the opposite. Quantum geometry gives rise to new discrete
structures at the Planck scale that modify the classical theory in
such a way that, at least in simple models, space-like
singularities of general relativity are resolved. However, since
the emphasis is on background independence and non-perturbative
methods, a priori it is not clear whether the theory also has a
rich semi-classical sector. Do the novel dynamical corrections
unleashed by the underlying quantum geometry naturally fade away
at macroscopic distances or do they have unforeseen implications
that prevent the theory from reproducing general relativity at
large scales? Some of such unforeseen problems are discussed in,
e.g.,  \cite{gu,aps2}.

\end{itemize}

Next, the dual role of the space-time metric also implies that
space-time itself ends when the gravitational field becomes
infinite. This is in striking contrast with Minkowskian physics,
where singularity of one specific field has no bearing at all on
the space-time structure or on the rest of physics. In general
relativity singularities of the gravitational field represent an
\emph{absolute boundary} of space-time where \emph{all of physics}
comes to a halt.

Now, it is widely believed that the prediction of a singularity
---such as the big-bang--- in classical general relativity is
primarily a signal that the theory has been pushed beyond the
domain of its validity and cannot therefore be trusted. One needs
a quantum theory of gravity to analyze true physics. This
expectation immediately leads to a host of questions which have
been with us for several decades now:\\
\begin{itemize}
\item How close to the big-bang does a smooth space-time of GR
make sense?  Inflationary scenarios, for example, are based on a
space-time continuum. Can one show from some first principles that
this is a safe approximation already at the onset of inflation?

\item Is the Big-Bang singularity naturally resolved by quantum
gravity? This possibility led to the development of the field of
quantum cosmology in the late 1960s. The basic idea can be
illustrated using an analogy to the theory of the hydrogen atom.
In classical electrodynamics the ground state energy of this
system is unbounded below. Furthermore, because an accelerating
electron radiates continuously, it must fall into the proton
making the atom extremely unstable. Quantum physics intervenes
and, thanks to a non-zero Planck's constant, the ground state
energy is lifted to a finite value, $-me^4/2\hbar^2 \approx -
13.6{\rm ev}$. The hope was that a similar mechanism would resolve
the big-bang and big crunch singularities in simple cosmological
models. However, as we will see in some detail in section
\ref{s2}, the \WDW quantization did not realize this hope. Can the
quantum nature of geometry underlying LQG make a fundamental
difference to this status-quo?

\item Is a new principle/ boundary condition at the big bang or
the big crunch essential to provide a deterministic evolution? The
most well known example of such a boundary condition is the `no
boundary proposal' of Hartle and Hawking \cite{hh2}. Or, do
quantum Einstein equations suffice by themselves even at the
classical singularities?

\item Do quantum dynamical equations remain well-behaved even at
these singularities? If so, do they continue to provide a
deterministic evolution? The idea that there was a pre-big-bang
branch to our universe has been advocated by several approaches,
most notably by the pre-big-bang scenario in string theory
\cite{pbb1} and ekpyrotic and cyclic models inspired by brane
world ideas \cite{ekp1,ekp2}. However, in such perturbative
treatments there is always a smooth continuum in the background
and hence the dynamical equations break down at the singularity.
Consequently, the pre-big-bang branch is not joined to the current
post-big-bang branch by a deterministic evolution. The hope has
been that non-perturbative effects would remedy this situation in
the future. LQG, on the other hand, does not have a continuum
geometry in the background and the treatment is non-perturbative.
So, one is led to ask: Can the LQC quantum Einstein's equation
provide a deterministic evolution across the big bang and the
big-crunch?

\item  If there is a deterministic evolution, what is on the
`other side'? Is there just a quantum foam from which the current
post-big-bang-branch is born, say a `Planck time after the
putative big-bang'? Or, was there another classical universe as in
the pre-big-bang and cyclic scenarios, joined to ours by
deterministic equations?
\end{itemize}

\emph{The goal of this article is show that such questions can be
answered satisfactorily in the simplest LQC models}. The first
significant results appeared already five years ago in the
pioneering work of Bojowald \cite{mb1}. Since then there has been
a steady stream of papers by a dozen or so researchers and the
field achieved a new degree of maturity over the last year.
However, as discussed in section \ref{s4} this is not the end of
the story because in LQC one first symmetry reduces the classical
theory and then quantizes it. Nonetheless, since the quantization
procedure mimics that of full LQG, the LQC answers provide not
only the much needed intuition but also a strategy for answering
the larger questions.

The article is organized as follows. Section \ref{s2} introduces
the reader to quantum cosmology and reviews the \WDW theory. While
one can answer some of the questions posed above, the \WDW theory
fails to resolve the big bang and the big crunch singularities.
The situation is drastically different in LQC where not only are
the singularities resolved but most of the questions have
physically desired answers. The key ideas and the structure of LQC
are summarized in section \ref{s3}, which also explains the sense
in which it `descends' form full LQG and emphasizes the main
differences between LQC the \WDW theory. For definiteness, I will
use the k=1 FRW models coupled to a massless scalar field
\cite{warsaw,apsv} to illustrate in some detail both the
difficulties discussed above and the way they can be handled. For
results on the k=0 model, see \cite{aps1,aps3}, for a first
discussion of the k=1 and of the anisotropic models, see
\cite{kv,kasner} and for a detailed review of the developments in
LQC till 2005, see \cite{mbrev}. I conclude in section \ref{s4}
with a brief summary of these other developments and a discussion
of the many challenges that still remain.

\section{Quantum Cosmology: The \WDW Theory}
\label{s2}

Both on the mathematical and phenomenological fronts, most of the
work to date in classical cosmology involves spatially homogeneous
models and perturbations thereof. In the late 60's DeWitt
\cite{bd} and Misner \cite{cm} began investigations in
\emph{quantum} cosmology with a similar viewpoint: first reduce
general relativity by imposing homogeneity and then quantize the
resulting system. Since this system has only a finite number of
degrees of freedom, the key field-theoretic difficulties are
bypassed. However, since there is no classical space-time in the
background, most of the conceptual problems of full quantum
gravity still remain. Therefore one can hope that resolution of
these problems in this technically simpler context would provide
valuable insights for the full theory. In particular, as mentioned
above, an initial expectation was that quantum effects
---particularly the uncertainty principle--- would intervene and
resolve the big bang and the big crunch singularities. While
subsequent work in the seventies and eighties did shed light on
several conceptual issues, this specific expectation was not met.
I will begin by briefly explaining the strategy used in these
analyses and point out its limitation. This summary will also
serve to bring out the new elements of LQC.

In this early work, the configuration space consisted of positive
definite 3-metrics. In the simplest, spatially homogeneous,
isotropic context, this metric is completely determined by the
scale factor $a$. Since $a$ is restricted to be positive, one
generally introduced a new variable $\a = \ln a$ and, as in
quantum mechanics, considered wave functions $\Psi(\a,\phi) \in
L^2 (\R^2, \dd \a\dd \phi)$, where $\phi$ denotes possible matter
fields. As in quantum mechanics, the basic operators were defined
via $\hat{\a}\,\Psi = \a \Psi,\,\, \hat{p}_\a\,\Psi = -i\hbar\,
(\p \Psi/\p \a),\,\, \hat\phi\,\Psi = \phi \Psi$, and
$\hat{p}_\phi\, \Psi = -i\hbar\,(\p\Psi/\p \phi)$. Quantization of
the classical Hamiltonian constraint,
\be \fs{2\pi G}{3a^3}\, p_\a^2 + \fs{3\pi^2}{2G}\, e^{\a} = H_\phi
\, , \ee
where $H_\phi$ is the matter Hamiltonian, then led to a
differential equation, called the \WDW equation. If the matter
field is a zero rest mass scalar field
---the case I will focus on for simplicity--- the \WDW equation
becomes
\be \label{wdwhc} \fs{4\pi}{3G}\, {\partial^2_\a \Psi} -
\fs{3\pi^2}{G\hbar^2} \, e^{4\a}\, \Psi = \partial^2_\phi \Psi \ee
(For details, see, e.g., \cite{ck}). Physical quantum states are
solutions to this equation. To extract physics, one has to
introduce an inner product and suitable observables on the space
of these states. The task of finding these observables is not
entirely straightforward because they must \emph{preserve the
space of solutions to (\ref{wdwhc})}, i.e., they must be Dirac
observables. Using these observables, one can then ask if
classical singularities are resolved in quantum theory. The older
\WDW literature does not appear to analyze this issue in any
detail since its focus was on the WKB approximation which,
unfortunately, becomes unreliable near the singularity. However,
they were analyzed in some detail recently (see e.g.
\cite{aps1,aps2,aps3,apsv}). I will now provide a summary of the
main results.

\begin{figure}[]
  \begin{center}
$a)$\hspace{8cm}$b)$
    \includegraphics[width=3.2in,angle=0]{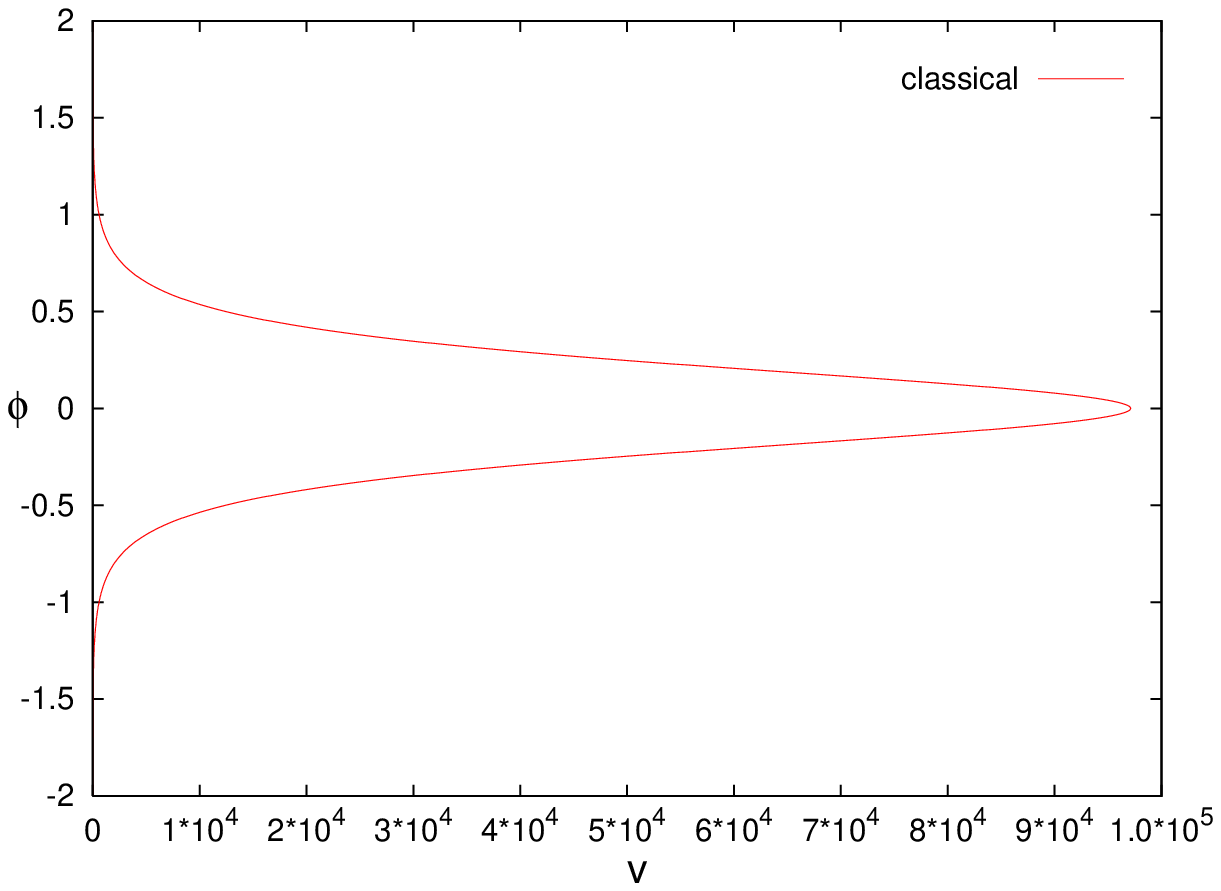}
\includegraphics[width=3.2in,angle=0]{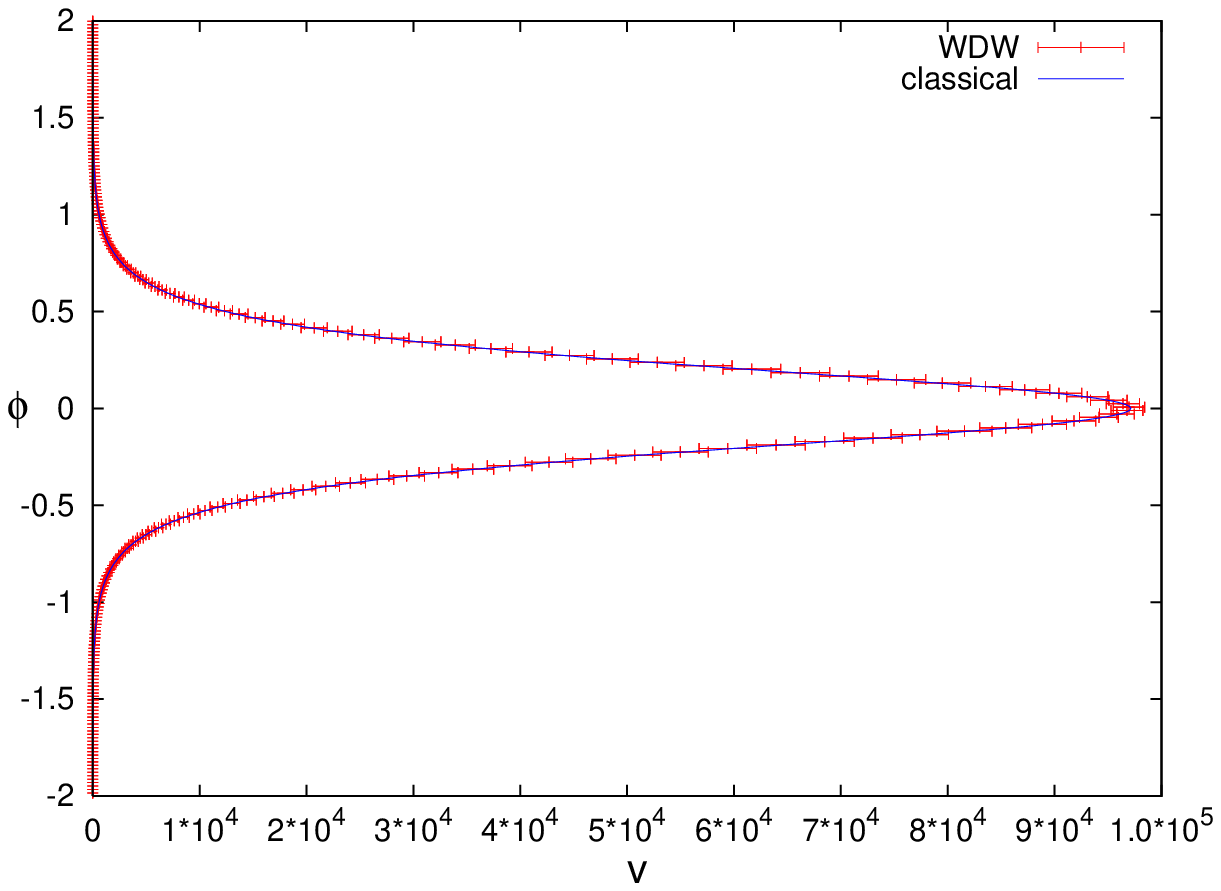}
\caption{$a)$ Classical solutions. Since $p_\phi$ is a constant of
motion, a classical trajectory can be plotted in the $v$-$\phi$
plane, where $v$ is essentially the volume in Planck units (see Eq
(\ref{v} )). $b)$ Expectation values (and dispersions) of
${|\hat{v}|_{\phi}}$ for the WDW wave function and comparison with
the classical trajectory. The \WDW wave function follows the
classical trajectory into the big-bang and big-crunch singularities.
(In this simulation, the parameters were: $p_{\phi}^{\star} = 5000
$, and $\Delta p_\phi/p_\phi^{\star} = 0.02$.) }
    \label{fig:wdw1}
  \end{center}
\end{figure}

Each classical trajectory begins with the big bang, undergoes a
subsequent expansion and a recollapse and finally contracts into
the big crunch singularity (see Fig. \ref{fig:wdw1} a). Since
there is no potential, the matter momentum $p_\phi$ is a constant
of motion. Consequently, $\phi$ is a monotonic function on each
classical trajectory. Therefore, it can be thought of as `internal
time' in the classical theory. The form of the \WDW equation
(\ref{wdwhc}) suggests that the same is true in quantum theory.
Thus the \WDW equation can be regarded as an `evolution equation'
with respect to this \emph{emergent time}, $\phi$. Next,
$\hat{p}_\phi = -i\hbar \partial_\phi$ is a Dirac observable in
the quantum theory. Given any value $\phi_o$ of $\phi$, one can
also define a second Dirac observable, $\hat{V}|_{\phi_o}$,
corresponding to the volume of the universe at any given `time'
$\phi_o$ \cite{aps1,aps2,aps3,apsv}. The pair $(\hat{p}_\phi,
\hat{V}_{\phi_o})$ constitutes a complete set of Dirac
observables. Finally, the requirement that these observables be
self-adjoint fixes the physical inner product (which can also be
determined by the more precise group averaging technique
\cite{dm}). Thus, the first set of questions raised in section
\ref{s1} can be answered satisfactorily in the \WDW theory.

Now to answer the remaining questions about singularity, one can
proceed as follows. Consider a point $(v^\star,\, \phi^\star)$ on
a classical trajectory with $p_\phi = p_\phi^\star$, where the
universe is large and the curvature is low compared to the Planck
scale. At `time' $\phi =\phi^\star$, one can construct an initial
\emph{quantum} state which is peaked at values $p_\phi^\star,
v^\star$ of the Dirac observables $\hat{p}_\phi$ and
$\hat{V}_{\phi_o}$. One can then evolve it using the \WDW
equation.%
\footnote{Technically, physical states satisfy the  `positive
frequency' square root $i\partial_\phi \Psi = -
\sqrt{\underline\Theta} \Psi$ of (\ref{wdwhc}), where the positive
definite self-adjoint operator $\underline\Theta$ is the negative
of the operator on the right side of (\ref{wdwhc}). One uses this
`first order in time' equation to `evolve' the wave function. For
details, see \cite{aps2,aps3,apsv}.}
The solutions can be expressed in terms of modified Bessel
functions. A resulting plot of the expectation values and
dispersions of $\hat{V}_\phi$ as a function of $\phi$ is shown in
Fig. \ref{fig:wdw1} b. The good news is that, when the universe is
large, the quantum state remains sharply peaked at the classical
trajectory, showing that the theory has a good semi-classical
limit. The bad news is that the quantum state continues to be
peaked at this trajectory even as the classical trajectory hits
the big bang and the big crunch singularities! Thus, although the
\WDW theory is background independent and non-perturbative, its
results are similar to that of background dependent, perturbative
treatments: Good semi-classical limit but no improvement on the
classical short distance pathologies.\bigskip

\textbf{Remark:} In the classical solution, the scalar field
$\phi$ tends to $-\infty$ at the big bang and to $\infty$ at the
big crunch. Therefore, the \WDW evolution is unitary on the entire
real line of the `emergent time'. However, as we just saw, even
semi-classical states simply track the classical solution and
therefore to the extent classical solutions are singular
--singularities are reached in finite \emph{proper time}-- so are
the quantum solutions. Thus, while the emergent time $\phi$ is
useful in enabling us to think of `evolution', unitary evolution
over its entire range does not guarantee non-singular behavior.
Additional considerations, such as the relation between the
emergent time and proper time in semi-classical settings are
essential.

\section{Loop Quantum Cosmology: The gravitational Sector}
\label{s3}
\subsection{Classical Theory}
\label{s3.1}

Since readers may not be familiar with the Hamiltonian description
used in LQC, I will summarize its key elements before proceeding
with quantization.

Let us consider space-time manifolds which have the form $M\times
\R$, where $M$ has the topology of a 3-sphere, $\S^3$. As Misner
showed in the late 60's \cite{cm}, one can identify $M$ with the
symmetry group $\SU(2)$ (which ensures spatial homogeneity and
isotropy) and endow it with a fixed fiducial basis of 1-forms
$\w_a^i$ and vectors $\e^a_i$. The resulting fiducial metric is
\be \q_{ab}:= \w_a^i\, \w_b^j\, k_{ij}, \quad\quad  \hbox{\rm
$k_{ij}$:\,\,\,  the Cartan-Killing metric on $\su(2)$.} \ee
$\q_{ab}$ turns out to be the metric of the round 3-sphere with
radius $a_o=2$ (rather than $a_o=1$). The volume of $M$ w.r.t.
this fiducial metric $\q_{ab}$ is $V_o = 2\pi^2a_o^3 = 16\pi^2$
and the scalar curvature is ${}^o\!R = 6/a_o^2 = 3/2$. We shall
set $\lo := V_o^{1/3}$.

In LQG, the dynamical variables are $\SU(2)$ (gravitational)
spin-connections $A_a^i$. Their conjugate momenta $E^a_i$ are the
analogs of electric fields in Yang-Mills theory. However, since
the `internal' group $\SU(2)$ is now tied with the group $\SO(3)$
of rotations in the tangent space at any point of $M$, $E^a_i$ now
acquire a direct geometric meaning: they represent spatial,
orthonormal triads (with density weight 1). In the present
isotropic, homogeneous setting, these canonically conjugate fields
are parameterized by constants $c$ and $p$ respectively:
\be \label{ps} A_a^i = c\, \lo^{-1}\,\, \w_a^i, \quad{\rm and}
\quad E^a_i = p\, \lo^{-2}\,\sqrt{\q}\,\, \e^a_i\,  \ee
which are functions of time. The factors of $\l_o$ facilitate
comparison with the spatially flat, $k$=0 case
\cite{aps1,aps2,aps3}. In both cases, $c$ is dimensionless while
$p$ has dimensions of area. In terms of geometrodynamical
variables, $p$ determines the scale factor $a$ (which determines
the spatial metric) and $c$, the canonically conjugate momentum,
or, $\dot{a}$ (which determines the extrinsic curvature). More
precisely, at the point $(c,p)$ of the phase space, the physical
3-metric $q_{ab}$ and the extrinsic curvature $K_{ab}$ are given
by:
\be q_{ab} = |p|\,\, \lo^{-2} \, \q_{ab}\quad {\rm and}\quad
\g\,K_{ab} = (c-\textstyle{\frac{\lo}{2}}) \,\,
|p|^{\fs{1}{2}}\,\, \lo^{-2} \, \q_{ab} \ee
The corresponding physical volume of $M$ is $V = |p|^{\fs{3}{2}}$.
The scale factor ${a}$ associated with a physical metric $q_{ab}$
is generally expressed via $q_{ab} = {a}^2 \,
{}^o\!\underbar{q}_{ab}$ where ${}^o\! \underbar{q}_{ab}$ is the
\emph{unit} 3-sphere metric. Then, the scale factor is related to
$p$ via $|p| =  a^2 \lo^2 / 4$. $p$ can take both positive and
negative values, the change in sign corresponds to a flip in the
orientation of the triad which leaves the physical metric $q_{ab}$
invariant.

Using the fact that $A_a^i$ and $E^a_i$ are canonically conjugate,
it is easy to calculate the fundamental Poisson bracket between
$c$ and $p$:
\be \label{pb} \{c,\, p\} = \f{8\pi G\g}{3}   \ee
where $\g$ is the so called `Barbero-Immirzi parameter' which
labels quantization ambiguity of LQG. (Black hole entropy
considerations show that $\g \approx 0.24$.) Finally, using the
fact that the co-triad $\w_a^i$  satisfies the Cartan identity
\be \label{cartan} \dd \w^k + \frac{1}{2}\, \ep_{ij}{}^k
\w^i\wedge \w^j =0\, , \ee
it is straightforward to calculate the field strength $F_{ab}^k$
of the connection $A_a^i$ on $M$:
\be \label{F1} F_{ab}^k = \lo^{-2}\, \big[ c^2 - c\,\lo \big]
\epsilon_{ij}{}^k\, \w_a^j\, \w_b^k\, . \ee

Because of our choice of parametrization of $A_a^i$ and $E^a_i$,
the only non-trivial constraint is the Hamiltonian one. Its
gravitational part reduces to \cite{apsv}:
\ba C_{\mathrm{grav}} &=& -\f{1}{\g^2}\,\, \int_{M} \dd^3 x\,
\ep^{ij}{}_{k} \, e^{-1}\, E^{a}_{i}E^{b}_{j}\, \, \bigg[F_{ab}^k
\,-\, \big(\f{1+\g^2}{4}\big)\, {}^o\!\epsilon_{ab}{}^c\,\,
\w_c^k\bigg]
\label{ham2}\\
&=& - \f{6\sqrt{p}}{\g^2}\, \bigg[ \big(c - \f{\lo}{2}\,\big)^2 +
\f{\g^2\lo^2}{4}\, \label{ham3}\bigg]\ea

\subsection{Quantum Kinematics}
\label{s2.2}

Since the phase space is now parameterized by $(c,p; \phi,
p_\phi)$, as in the \WDW theory one's first impulse would be to
follow standard quantum mechanics. In the \WDW case this strategy
was natural because one did not have access to full quantum
geometrodynamics to take guidance from. In LQC it is not: a more
appropriate strategy would be to follow the procedures used in
full LQG. One's first reaction may be: How can this make any
difference? After all the system has only a finite number of
degrees of freedom and von Neumann's theorem assures us that,
under appropriate assumptions, the resulting quantum mechanics is
unique. The only remaining freedom appears to be in the
factor-ordering of the Hamiltonian constraint and this is
generally insufficient to lead to qualitatively different
predictions. It turns out however that if one mimics full LQG, one
naturally violates a key assumption of von Neumann's uniqueness
theorem and, in spite of the presence of only a finite number of
degrees of freedom, one is led to \emph{new quantum mechanics}.

To explain this point, let me begin with a quick summary of the
relevant structure of full LQG. Since novel elements of interest
to us appear in the gravitational sector, in the remainder of this
subsection I will ignore matter. To maintain gauge covariance,
configuration variables are taken to be the `Wilson lines' or the
holonomies $h_e := {\cal P}\, \exp -\int_e A $ along arbitrary
edges $e$, determined by the $\SU(2)$ connection $A_a^i$.
Geometrically, being 1-forms, connections are naturally `smeared'
along 1-dimensional curves. The conjugate momenta are
density-weighted triads $E^a_i$ which (being duals of a triplet of
2-forms) are naturally smeared along 2-surfaces $S$ with test
fields $f_i$: $E(f) = \int_S {}^\star{E}_i f^i\, $. These are
referred to as triad fluxes. Note that these two sets of phase
space functions $(h_e(A),\, E(S,f))$ have been constructed without
any reference to a background metric or any other background
field. They generate an algebra, rather analogous to that
generated by functions $\exp i\lambda q$ and $p$ on the phase
space of a particle. Our job is to find a representation of this
`holonomy-flux' algebra which preserves background independence of
the theory. The surprising result is that the representation is
unique \cite{lost}! More precisely, the algebra admits a unique
(internal, i.e. $\SU(2)$ gauge and) diffeomorphism invariant state
and the representation is generated by the action
of the algebra on this state.%
\footnote{In the framework of algebraic quantum field theory, a
$\star$-algebra $\mathfrak{a}$ of operators is first constructed
abstractly, without reference to a Hilbert space. A state $f$ is a
positive linear functional on the algebra, i.e. a linear
functional which satisfies $f(A^\star A) \ge 0$ for all $A \in
\mathfrak{a}$. Such a state determines, via a standard
construction due to Gel'fand, Naimark and Segal, a $\star$
representation of $\mathfrak{a}$ on a Hilbert space, in which the
abstract state $f$ is now represented by a vector $\ket{\Psi_f}$
in the Hilbert space and its action by the expectation value
functional: $f(A) = \ip{\Psi_f}{A\,\Psi_f}$. The vector $\Psi_f$
is `cyclic' in the sense that a dense subspace of the full Hilbert
space is generated by repeated action of operators in
$\mathfrak{a}$ on $\Psi_f$. Finally, note that the uniqueness
result refers to the holonomy-flux algebra whose elements are not
diffeomorphism invariant. An algebra of diffeomorphism invariant
operators would admit many diffeomorphism invariant states.}
While the uniqueness result is rather recent, the representation
itself was constructed over a decade ago and is well understood
\cite{alrev,crbook,ttbook}. For our purposes a key feature of this
representation is the following: While there are well-defined
operators $\h{h}_e$ corresponding to holonomies, \emph{there is no
operator corresponding to the connection itself}.

In quantum cosmology one follows the same overall procedure. In
the k=1 model now under consideration, holonomies along integral
curves of the left invariant vector fields $\e^a_i$ already
suffice to separate homogeneous, isotropic connections. Let us
denote by $\lambda\lo$ the directed length (w.r.t. the fiducial
metric $\q_{ab}$) of the curve along $\e^a_i$ (so that $\lambda$
is positive if the curve is oriented in the direction of $\e^a_i$
and negative if has the opposite orientation). Then the holonomy
along the edge of length $\lambda\lo$ in the $k$th direction is
given by
\be h^k_{(\lambda)}(c) = \cos (\lambda c/2)\, \mathbb{I} + 2\sin
(\lambda c/2)\, \tau^k\ee
where $\mathbb{I}$ is the identity $2\times 2$ matrix. (The holonomy
is of course independent of the background metric $\q_{ab}$.) The
functions of $c$ which enter as coefficients are `almost periodic'
functions of $c$, i.e. are of the form $N_{(\lambda)}(c) := \exp
i\lambda (c/2)$. In the quantum theory, then, we are led to a
representation in which operators $\h{h}^k_{(\lambda)},\,
\h{N}_{(\lambda)}$ and $\h{p}$ are well-defined, but there is
\emph{no} operator corresponding to the connection component $c$.

At first this seems surprising because our experience with quantum
mechanics suggests that one should be able to obtain the operator
analog of $c$ by differentiating $\h{N}_{(\lambda)}$ with respect
to the parameter $\lambda$. However, in the representation of the
basic quantum algebra that descends to LQC from full LQG, although
the $\h{N}_{(\lambda)}$ provide a 1-parameter group of unitary
transformations, the group fails to be weakly continuous in
$\lambda$. Therefore one can not differentiate and obtain the
operator analog of $c$. In quantum mechanics, this would be
analogous to having well-defined (Weyl) operators corresponding to
the classical functions $\exp i\lambda x$ but no operator $\h{x}$
corresponding to $x$ itself. This violates one of the assumptions
of the von-Neumann uniqueness theorem. New representations then
become available which are \emph{inequivalent} to the standard
Schr\"odinger one. In quantum mechanics, these representations are
not of direct physical interest because we need the operator
$\h{x}$. In LQC, on the other hand, full LQG naturally leads us to
a new representation. \emph{This theory is inequivalent to the
\WDW type theory already at a kinematical level.}

The structure of this theory can be summarized as follows. Since
$\h{p}$ is a self-adjoint operator, the gravitational part $\Hkg$
of the kinematic Hilbert space can be described in terms of its
eigenbasis. A general state $|\Psi\rangle$ has the form
\be \label{state} |\Psi\rangle = \sum_i \Psi_i \ket{p_i}\quad {\rm
with} \quad \sum_i |\Psi_i|^2 < \infty\, , \ee
where $i$ ranges over a countable set and $\ket{p_i}$ is an
orthonormal basis:
\be\label{ip} \ip{p_i}{p_j} = \delta_{ij}\, .  \ee
While the general form of these equations may seem familiar from
quantum mechanics, there are key differences: the state is a
\emph{countable} sum of eigenstates of $\h{p}$ rather than a
direct integral and the right side of (\ref{ip}) is a Kronecker
delta rather than the Dirac delta distribution.%
\footnote{Thus, in contrast to the Schr\"odinger Hilbert space
$L^2(\R, dc)$, the kinematical LQC Hilbert space $\Hkg$ is
non-separable. However, as we will see, quantum dynamics is
governed by a discrete evolution equation. This leads to
superselection sectors so that the final physical Hilbert spaces
are all separable. For details, see \cite{aps2,apsv}.}
Consequently, the intersection of $\Hkg$ and $L^2(\R, \dd p)$
consists only of the zero function. However, the action of the
basic operators on wave functions $\Psi(p) := \langle p|\Psi
\rangle$ is the expected one:
\be \h{p} \Psi(p) = p \Psi(p) \quad {\rm and} \quad
\h{N}_{(\lambda)} \Psi (p) = \Psi(p+\lambda) \ee
The forms of these operators are the same as in ordinary quantum
mechanics. However, in the new Hilbert space $\h{N}_{(\lambda)}
\equiv \widehat{\exp i \lambda (c/2)}$ fails to be (weakly)
continuous in $\lambda$ because $\Psi(p+\lambda)$ is
\emph{orthogonal to} $\psi(p+{\lambda^\prime})$ for all
$\lambda\not= \lambda^\prime$. Consequently, one can not take a
derivative of $\h{N}_{(\lambda)}$ w.r.t. $\lambda$ and define an
operator $\h{c}$. Finally, let us note a fact that will be useful
in section \ref{s3.3}: from the discussion of the classical theory
of section \ref{s3.1} it follows that the physical volume
operator of $M$ is given by: $\h{V} \,=\, |\hat{p}|^{{3}/{2}}$. \\

\textbf{Remark:} Kinematics of LQC provides a useful, broad-brush
picture of the kinematics of full LQG (for details, see
\cite{alrev,crbook,ttbook}). We saw that the LQC kinematical
Hilbert space is \emph{not} the space $L^2(\R, \dd c)$ that one
would use in the \WDW theory. But it can in fact be written as the
space of square integrable functions, not on the classical
configuration space $\R$, but on a certain completion,
$\bar\R_{\rm Bohr}$ thereof, called the Bohr compactification of
the real line. This is an Abelian group with a natural Haar
measure. The kinematical Hilbert space can be written as $\Hkg
= L^2(\bar\R_{\rm Bohr}, \dd \mu_{\rm Haar})$.%
\footnote{ The space $\bar\R_{\rm Bohr}$ was introduced by the
mathematician Harold Bohr, Niels's brother, in his theory of
almost periodic functions. In the connection representation,
elements of $\Hkg$ are of the form $\Psi(c) = \sum_i \alpha_i\,
\exp{i\lambda_i c}$ where $\alpha_i$ are complex numbers,
$\lambda_i$ real and the sum extends over a countable set. These
$\Psi(c)$ are almost periodic functions of $c$, where the word
`almost' refers to the fact that $\lambda_i$ are arbitrary real
numbers. The scalar product given by (\ref{ip}) now becomes:
$\ip{\Psi_1}{\Psi_2} = \lim_{L\rightarrow \infty}\,
(1/2L)\,\int_{-L}^{L} \bar\Psi_1(c)\,\Psi_2(c)\, dc$.}
Thus, one way to understand the difference between the
Schr\"odinger and the new quantum mechanics is to realize that
while in the Schr\"odinger theory the `quantum configuration
space' continues to be the classical configuration space, $\R$, in
the new theory it is replaced by a certain completion $\bar\R_{\rm
Bohr}$ thereof. The same phenomenon occurs in full LQG: While the
classical configuration space $\A$ is the space of smooth $\SU(2)$
connections on a 3-manifold $\Sigma$, the quantum configuration
space $\Ab$ is a certain completion thereof, containing also
`generalized connections' which can be arbitrarily discontinuous.
The Haar measure on $\SU(2)$ induces a natural, faithful, regular
Borel measure $\dd\mu_o$ on $\Ab$. The kinematical Hilbert space
of LQG is $L^2(\Ab, \dd \mu_o)$. Thus the completion $\bar\R_{\rm
Bohr}$ of LQC is replaced by $\Ab$ and the measure $ \mu_{\rm
Haar}$ by the measure $\mu_o$. Technically both completions $\Ab$
and $\bar\R_{\rm Bohr}$ arise as the `Gel'fand spectrum' of the
holonomy algebras in the two theories.

Next, let us consider the algebraic approach. The $\star$-algebra
generated by $\h{N}_\lambda, \h{p}$ is the LQC analog of the
holonomy flux algebra of full LQG. The analog of the unique LQG
diffeomorphism state on this $\star$-algebra is the
positive-linear functional $f$ which has the following action on
the basic operators: $f(\h{N}_\lambda)= \delta_{\lambda, 0}$ and
$f(p)=0$. As in full LQG, the Gel'fand, Naimark, Segal
construction then leads us to the kinematical Hilbert space
$\Hkg$. In this space, the abstract state is represented by the
concrete vector $\ket{p=0}$, or equivalently $\Psi(c) = 1$, in
which the triad vanishes and, heuristically, the connection is
completely undetermined. As in LQG, the full Hilbert space is
generated by a repeated action of the algebra on this state. A
natural basis in the LQG kinematical Hilbert space is given by the
so-called spin network functions. These are analogous to the
states $N_\lambda(c)$ in the connection representation of LQC.
Just as $N_\lambda(c)$ diagonalize operators $\hat{p}$,
(appropriately chosen) spin networks diagonalize triads and hence
geometric operators. This is why a firm grasp of the LQC quantum
kinematics can provide good intuition for full LQG.

Finally, a word of caution: the analogy does fail in a few but
important respects because of the homogeneity assumption of LQC.
So, it should be pursued as a guideline to aid intuition rather
than a literal dictionary.

\subsection{Quantum Dynamics}
\label{s3.3}

To describe quantum dynamics, we have to first introduce a
well-defined operator on $\Hkg$ representing the Hamiltonian
constraint $C_{\rm grav}$. Since there is no operator corresponding
to $c$ itself, we cannot directly use the final expression
(\ref{ham3}). Rather, we return to the form (\ref{ham2}) from full
general relativity and promote it to an operator. This procedure
also serves to bring LQC closer to LQG. The general strategy is the
same as in ordinary quantum mechanics: One first expresses various
terms in (\ref{ham2}) in terms of `elementary variables', namely the
holonomies $h^k_{(\lambda)}$ and momenta $p$ which have direct
quantum analogs and then replaces them with operators
$\h{h}^k_{(\lambda)}$ and $\h{p}$. Quantization of the first term,
$\ep^{ij}{}_{k} \, e^{-1}\, E^{a}_{i}E^{b}_{j}\,$, can then be
carried out by directly following a procedure given by Thiemann in
the full theory \cite{tt,ttbook,abl}.

The second term is the field strength $F_{ab}{}^k$. As in gauge
theories, one can obtain ${F}_{ab}{}^k$ by first calculating the
holonomy around a closed loop in the a-b plane, multiplying it by
$\tau^k$, dividing by the area and taking the limit as the area
shrinks to zero. However, precisely because there is no connection
operator $\h{A}_a^i$ in LQG, the limit can not exist in quantum
theory.%
\footnote{In full LQG, one often defines a `Master Hamiltonian
constraint' only on diffeomorphism invariant states \cite{ttbook}.
On this restricted class of states, the limit of the constraint
operator as the area shrinks to zero is well defined. In LQC this
avenue is not easily available because the diffeomorphism freedom
is gauge fixed in the very beginning while parameterizing the
connection and triad by $c$ and $p$.}
However, the same feature that prevents $\hat{A}_a^i$ from
existing leads to quantum Riemannian geometry in LQG and in
particular implies that there is a smallest non-zero eigenvalue,
$\Delta$, of the area operator. Therefore the non-existence of the
limit is interpreted as telling us that the procedure of shrinking
the area of the loop to zero is inappropriate in quantum theory.
Rather, we should shrink the loop till its area equals $\Delta$.
Now, given a point $(c,p)$ of the classical phase space, a
`square' loop each of whose sides has length $\bar\lambda \l_o$
with respect to the fiducial metric $\q_{ab}$ has area
$\bar{\lambda}^2 |p|$ with respect to the \emph{physical} metric
$q_{ab} = |p| \l_o^{-2}\, \q_{ab}$. Therefore, to define
$\h{F}_{ab}{}^k$, the holonomy is evaluated around a `square' loop
whose edges are formed by the integral curves of right and left
invariant vector fields on $\S^3$, with $\bar\lambda$ given by
\be \mb^2 |p|\, =\, \Delta\, \equiv\, 2\sqrt{3}\, \pi \gamma\,\,
\lp^2\, , \ee
so that \emph{physical} area of the surface enclosed by the loop is
$\Delta$. (Because of homogeneity, the precise location of the loop
is irrelevant.) The resulting $\hat{F}_{ab}{}^k$, given by
\be \label{Fop} \h{F}_{ab}{}^k = \f{1}{\mb
\l_o^2}\,\big(\sin^2\mb(c - {\l_o}/{2})
-\sin^2(\mb{\l_o}/{2})\big)\, \epsilon_{ij}{}^k\, \w_a^i\,
\w_b^j,\ee
is well-defined but, as in full LQG, has a built in fundamental
non-locality.

It was clear from the beginning that something like this must
occur because there is no connection operator in LQG. Recall that
passage to quantum theory from the classical one \emph{always}
requires new physical inputs. Quantization of ${F}_{ab}{}^k$ is
well motivated in the representation of the holonomy-flux algebra
we are forced to use by the requirement of back ground
independence. Furthermore, one can show that $\hat{F}_{ab}{}^k$
does reduce, in a precise sense, to the standard local expression
of curvature $F_{ab}{}^k$ in the classical limit. (For details,
see \cite{aps3,apsv}.) Hence $\hat{F}_{ab}{}^k$ is a well-defined
quantization of $F_{ab}{}^k$. However, the non-locality makes a
crucial difference in the Planck regime. In particular, as we will
now see, it makes the gravitational part $\h{C}_{\rm grav}$ of the
constraint a \emph{difference} operator, rather than a
differential operator as in the Wheeler-DeWitt theory. As one
might expect, the `step size' is governed by the area gap
$\Delta$.

Using (\ref{Fop}) and the the operator corresponding to
$\ep^{ij}{}_{k} \, e^{-1}\, E^{a}_{i}E^{b}_{j}\,$ given by the
Thiemann procedure it is straightforward to construct the
gravitational part $\h{C}_{\rm grav}$ of the constraint operator
corresponding to (\ref{ham2}). It turns out \cite{aps3,apsv} that
its expression is simplest if one considers wave functions which are
naturally diagonal in the volume operator $\h{V}$ rather than in
$\h{p}$. (Mathematically, this is a rather trivial transformation
since $\h{V} = |\h{p}|^{3/2}$.) It is convenient to parameterize the
eigenvalues of the volume operator by $v$ such that:
\be \label{v} \h{V}\ket{v} =  (\f{8\pi\g}{6})^{\f{3}{2}}\,\,
\f{|v|}{K}\,\,\lp^3 \,\, \ket{v}\, \quad {\rm where} \quad K =
\f{2\sqrt{2}}{3\sqrt{3\sqrt{3}}}\, . \ee
Then, the gravitational part of the constraint is given by:
\ba \label{ham4} \h{C}_{\mathrm{grav}}\, \Psi(v) \,= &&e^{if\lo}
\,\,\sin\mb c\, \h{A} \sin\mb c\,\, e^{-i\lo f}\,\, \Psi(v)\nonumber\\
&-&\big[\sin^2\f{\mb\lo}{2} - \f{\mb^2\lo^2}{4} - \f{\lo^2}{9|K^2
v|^{\fs{2}{3}}}\big]\,\, \h{A}\, \Psi(v)\, \ea
Here, $f$ is a simple function of $v$,\, $f(v)= ({\rm sgn}\,
v/4)\,\, |v/K|^{2/3}$, and the operator $\h{A}$, given by
\be \label{A} \hat A \,\Psi(v)\, =\, - \f{27K}{4}
\,\sqrt{\f{8\pi}{6}} \, \f{\lp}{\gamma^{3/2}} \,|v| \,\,\, \big| |v
- 1| - |v + 1| \big|\, \,\, \Psi(v) ~, \ee
is also diagonal in $v$.

To summarize, if we mimic full LQG, we are naturally led to a theory
that is inequivalent to the \WDW theory even at the kinematical
level. Indeed the two Hilbert spaces have no non-zero element in
common. A key distinguishing feature of this theory is that while
the holonomies of the connection are well defined operators,
connections themselves are not. Consequently, in the definition of
the quantum Hamiltonian constraint (\ref{ham2}), we are led to
define the field strength through holonomies along loops which
enclose area $\Delta$, the smallest non-zero eigenvalue of the area
operator. The gravitational part of the resulting Hamiltonian
constraint is a \emph{difference} operator, rather than a
differential operator that would have resulted had we worked in a
Schr\"odinger type representation used in the \WDW theory. However,
in a well-defined sense it reduces to the \WDW differential operator
for large $v$ \cite{apsv}.

\subsection{The Full Hamiltonian Constraint and Main Results}
\label{s3.4}

As in the \WDW theory, we will now assume that the only matter
field is a massless scalar field (although it is straightforward
to allow additional matter fields). The form of the Hamiltonian
constraint is such that this field can serve as an internal clock
also in LQC and we can again use relational dynamics. This
simplifies the intermediate technical steps and makes the physical
meaning of results more intuitive.

To write the complete Hamiltonian constraint we also need the
matter part. For the massless scalar field, in the classical
theory it is given by:
\be C_{\rm matt} = 8\pi G\,\, |p|^{-\f{3}{2}}\, p_\phi^2 \ee
The non-trivial part in the passage to quantum theory is the
function $|p|^{-3/2}$. However, we can define this operator again
by using the method introduced by Thiemann in the full theory
\cite{tt,ttbook,abl}. The final result is \cite{aps3}:
\be \label{inversevol}  \widehat{{|p|^{-\f{3}{2}}}} \Psi(v) =
\left(\f{6}{8 \pi \gamma \lp^2}\right)^{3/2}\, B(v)\, \Psi(v) \ee
where
\be \label{B} B(v) = \left(\f{3}{2}\right)^3 \, K\,\, |v| \,
\bigg| |v + 1|^{1/3} - |v - 1|^{1/3} \bigg|^3\, . \ee
This operator is self-adjoint on $\Hkg$ and diagonal in the
eigenstates of the volume operator.

 We can express the total constraint
\be \label{hc1} \h{C}\, \Psi(v) = \left(\hat{C}_{\rm grav} +
\hat{C}_{\rm matt}\right)\, \Psi(v) = 0\, ,\ee
as follows:
\ba \label{hc3} \p^2_\phi \Psi(v,\phi) = &-& \Theta\, \Psi(v,\phi)
\nonumber\\
= &-& \Theta_o \Psi (v,\phi) + \f{\pi G}{2} [B(v)]^{-1} \Big[3K
(\sin^2(\f{\mb\lo}{2}) - \f{\mb^2\lo^2}{4})\, |v|\,
\nonumber\\
&-& \, \f{1}{3}\, \lo^2\g^2\,\,
\big|\f{v}{K}\big|^{\fs{1}{3}}\Big] \,\,\Big[ \big|\,
 |v-1|- |v+1|\,\big|\Big]\, \Psi(v,\phi)\, .\ea
Here, $\Theta_o$ is a difference operator,
\be \label{hc4} \Theta_o \Psi(v,\phi) =  - [B(v)]^{-1} \,
\left(C^+(v)\, \Psi(v+4,\phi) + C^o(v) \, \Psi(v,\phi) +C^-(v)\,
\Psi(v-4,\phi)\right)\, , \ee
where the coefficients $C^\pm(v)$ and $C^o(V)$ are given by:
\ba \label{C} C^+(v) &=& \nonumber \f{3\pi K G}{8} \, |v + 2|
\,\,\,
\big| |v + 1| - |v +3|  \big|  \\
C^-(v) &=& \nonumber C^+(v - 4) \\
C^o(v) &=& - C^+(v) - C^-(v) ~. \ea
For the k=0 (i.e., flat) FRW model the quantum constraint is
obtained by simply replacing $\Theta$ by $\Theta_o$ in (\ref{hc3})
\cite{aps3}. Thus, the $k$=0 quantum constraint has the same form
as in the $k$=1 case. As one would expect from the classical
expression (\ref{hc1}), the difference $\Theta-\Theta_o$ is
diagonal in the $v$-representation and vanishes when we set
$\lo=0$.

\begin{figure}[]
  \begin{center}
    $a)$\hspace{8cm}$b)$
    \includegraphics[width=3.2in,angle=0]{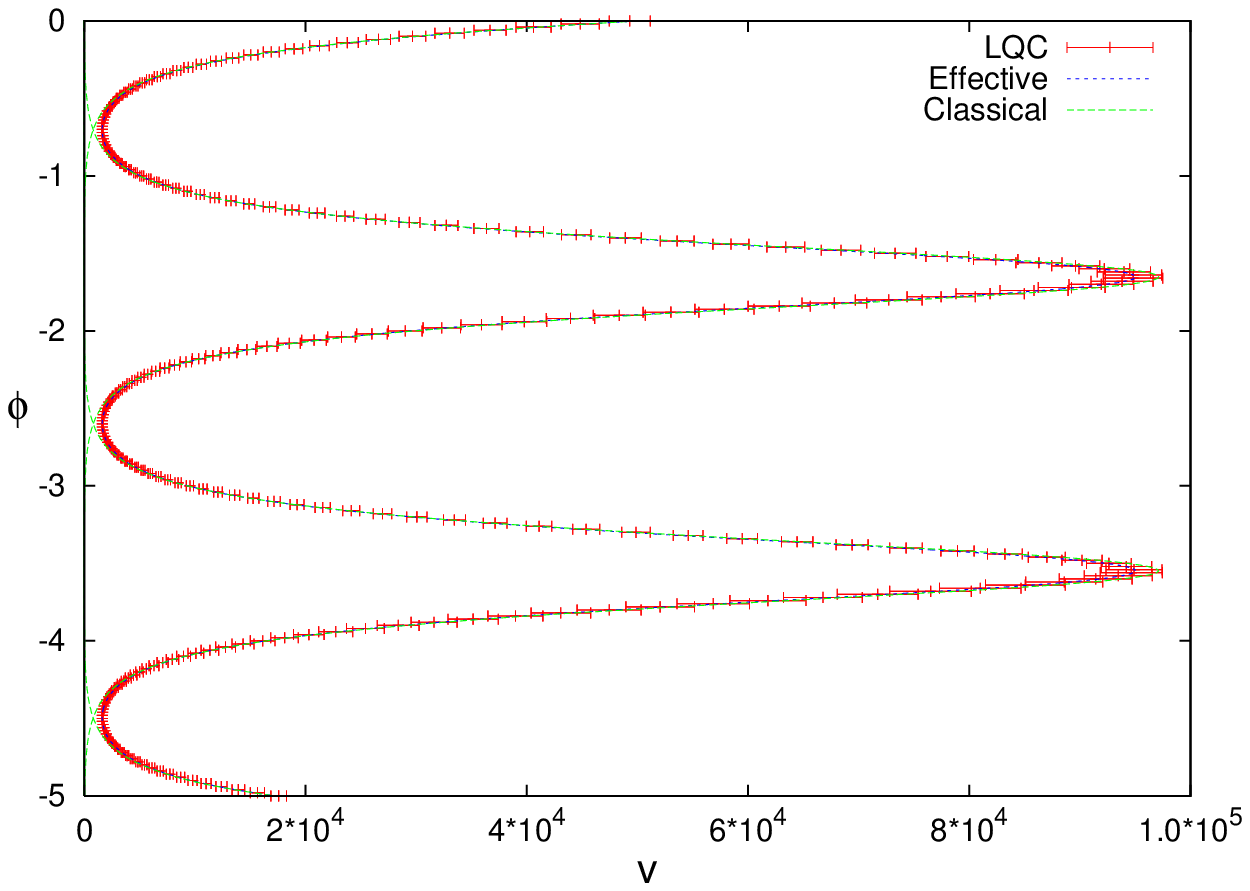}
    \includegraphics[width=3.2in,angle=0]{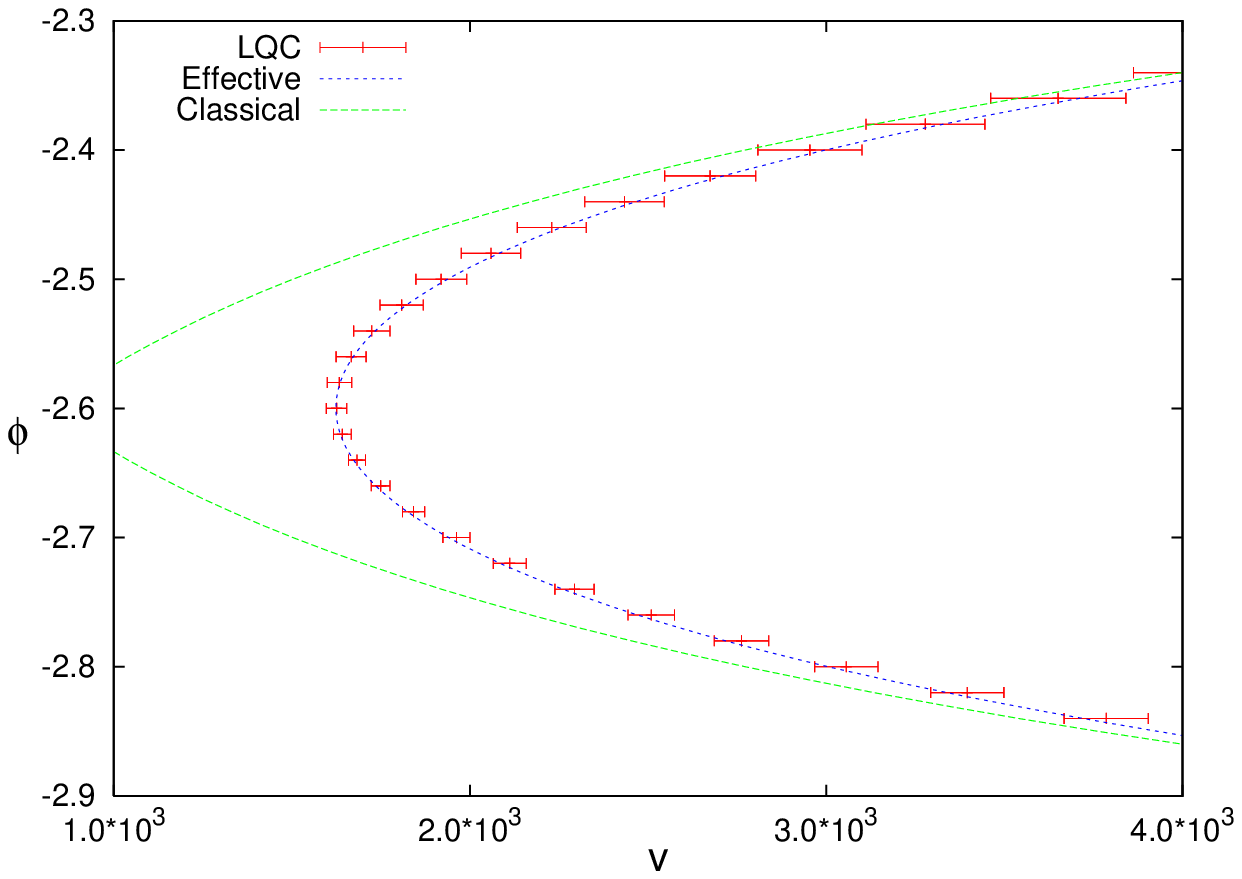}
\caption{In the LQC evolution the big bang and big crunch
singularities are replaced by quantum bounces. $a)$  Expectation
values and dispersion of $|\h{v}|_\phi$ are compared with the
classical trajectory and the trajectory from effective Friedmann
dynamics (see (\ref{eff})). The classical trajectory deviates
significantly from the quantum evolution at Planck scale and evolves
into singularities. The effective trajectory provides an excellent
approximation to quantum evolution at all scales. \,\, $b)$ Zoom on
the portion near the bounce point of comparison between the
expectation values and dispersion of ${\hat{v}|_{\phi}}$, the
classical trajectory and the solution to effective dynamics. At
large values of $|v|_\phi$ the classical trajectory approaches the
quantum evolution. In this simulation $p_\phi^\star = 5\times 10^3$,
$\Delta p_\phi/p_\phi^\star = 0.018$, and $v^\star = 5\times 10^4$.}
\label{fig:lqc}
\end{center}
\end{figure}

Let us continue with the k=1 case. The form of the Hamiltonian
constraint is similar to that of a massless Klein-Gordon field in
a static space-time, with a static potential. $\phi$ is the analog
of the static time coordinate and the difference operator
$\Theta$, of the spatial Laplace-type operator plus the static
potential. Hence, the scalar field $\phi$ can again be used as
`emergent time' in the quantum theory. As in the \WDW case,
$\h{p}_\phi$ and $\h{V}|_{\phi_o}$ provide a complete set of Dirac
observables. The physical inner product can again be fixed either
by requiring that these operators be self-adjoint or by the group
averaging method \cite{dm}. This provides us with the
\emph{physical sector} of the theory. As in the \WDW case, one can
construct states which are sharply peaked at given values
$p_\phi^\star$ and $v^\star$ of $\h{p}_\phi$ and
$\h{V}|_{\phi^\star}$ at an instant of `time' $\phi^\star$ and use
(\ref{hc3}) to `evolve' them. While the general structure of the
resulting LQC theory is thus analogous to that of the \WDW theory,
there is a dramatic difference the final results.

Numerical techniques play a vital role in this analysis. More
precisely, numerical evolutions have been performed for several
values of parameters and using two different techniques. They attest
to the robustness of results. While solutions mimic the \WDW
behavior when the universe is large, there is a radical departure in
the strong curvature region: \emph{the big bang and the big crunch
singularities are resolved and replaced by big-bounces.} The main
results can be summarized as follows. (For details, see
\cite{apsv}.)

\b  Consider a classical solution depicted in Fig.\ref{fig:wdw1} a
which evolves from the big-bang to the big crunch, reaching a
large maximum radius $a_{\rm max}$. Fix a point on this trajectory
where the universe has reached macroscopic size and consider a
semi-classical state peaked at this point. \emph{Such states
remain sharply peaked throughout the given `cycle', i.e., from the
quantum bounce near the classical big-bang to the quantum bounce
near the classical big-crunch.} Note that the notion of
semi-classicality used here is rather weak: these results hold
even for universes with $a_{\rm max} \approx 23 \lp$ and the
`sharply peaked' property improves greatly as $a_{\mathrm{max}}$
grows.

\b The trajectory defined by the expectation values of the Dirac
observable $\h{V}|_\phi$ in the full quantum theory is in good
agreement with the trajectory defined by the classical Friedmann
dynamics until the 4-d scalar curvature (the only independent
curvature invariant in isotropic, homogeneous models) attains the
value $\approx 13\pi/\lp^2$, or, equivalently, the energy density
of the scalar field becomes comparable to a critical energy
density $\rcr \approx 0.82\rho_{\rm Pl}$. \emph{Then the classical
trajectory deviates from the quantum evolution.} In the classical
solution, scalar curvature and the matter energy density keeps
increasing on further evolution, eventually leading to a big bang
(respectively, big crunch) singularity in the backward
(respectively, forward) evolution, when $v \rightarrow 0$. The
situation is very different with quantum evolution. Now the
universe bounces at $\rho \approx 0.82\rho_{\rm Pl}$, avoiding the
past (or the big bang) and future (or the big crunch)
singularities.

\b The volume of the universe takes its minimum value $V_{\rm
min}$ at the bounce point. $V_{\rm min}$ scales linearly with
$p_\phi$:
\footnote{Here and in what follows, numerical values are given in
the classical units G=c=1. In these units $p_\phi$ has the same
physical dimensions as $\hbar$ and the numerical value of $\hbar$
is $2.5\times 10^{-66}{\cm}^2. $}
\be V_{\rm min} = \big( \f{4\pi G\g^2 \Delta}{3}
\big)^{\f{1}{2}}\, p_\phi \,\, \approx (1.28\times 10^{-33}\,
{\cm})\,\, p_\phi\ee
Consequently, $V_{\rm min}$ can be \emph{much} larger than the
Planck size.  Consider for example a quantum state describing a
universe which attains a maximum radius of a megaparsec. Then the
quantum bounce occurs when the volume reaches the value $V_{\rm min}
\approx 5.7 \times 10^{16}\, {\cm}^3$, \emph{some $10^{115}$ times
the Planck volume.} Deviations from the classical behavior are
triggered when the density or curvature reaches the Planck scale,
even when the volume is large. Since $V_{\rm min}$ \emph{is} large
at the bounce point of macroscopic universes, the so-called `inverse
volume effects' ---i.e. the fact that $B(v) \not= 1/v$--- are
largely insignificant to the quantum dynamics of such universes.

\b After the quantum bounce the energy density of the universe
decreases and, when $\rho \ll \rmax$, the quantum evolution is
well-approximated by the classical trajectory. On subsequent
evolution, the universe recollapses both in classical and quantum
theory at the value $V=V_{\mathrm{max}}$ when energy density
reaches a minimum value $\rmin$.  $V_{\rm max}$ scales as the
3/2-power of $p_\phi$:
\be\label{Vmax}  V_{\rm max} = (16\pi G/3 \lo^2)^{\fs{3}{4}}\,
p_\phi^{\fs{3}{2}}\,\, \approx \,\,  0.6\, p_\phi^{\fs{3}{2}} \ee
Quantum corrections to the classical Friedmann formula $\rmin =
3/8\pi Ga^2_{\rm max}$ are of the order $O(\lp/a_{\rm max})^4$.
For a universe with $a_{\rm max} = 23\lp$, the correction is only
one part in $10^{-5}$. For universes which grow to macroscopic
sizes, classical general relativity is essentially exact near the
recollapse.

\b There is a Hamiltonian analog of the `effective action' framework
\cite{jw,apsv} which provides a systematic procedure to obtain
`effective equations' from quantum theory. While the classical
Friedmann equation is $(\dot{a}/a)^2 = ({8\pi G}/{3})\, (\rho-
3/8\pi G a^2)$, the effective Friedmann equation turns out to be
\be \label{eff} \left(\f{\dot{a}}{a} \right)^2 = \fs{8\pi G}{3}\,
(\rho-\rho_1(v))\,\, \left[\rho_2(v) - \rho/\rho_{\rm crit}\right]
\ee
where $\rho_1$ and $\rho_2$ are specific functions of $v$. The
term in the square bracket is the key quantum correction; away
from the Planck regime, $\rho_2 \approx 1$ and $\rho/\rho_{\rm
crit} \approx 0$. Bounces occur when $\dot{a}$ vanishes, i.e. at
the value of $v$ at which the matter density equals $\rho_1(v)$ or
$\rho_2(v)$. $\rho(v) = \rho_1(v)$ at the classical recollapse
while $\rho(v)=\rho_2(v)$ at the quantum bounce.%
\footnote{For $k$=0, i.e. open universes, the Friedmann equation
$(\dot{a}/a)^2 = (8\pi G/3)\, \rho$ is replaced just by
$(\dot{a}/a)^2 = (8\pi G/3)\, (\rho - \rcr)$. Since there is no
classical recollapse, there is a single pre-big-bang contracting
branch which is joined deterministically to the post-big-bang
expanding branch. For details see \cite{aps3}.}

\b For quantum states under discussion, the density $\rmax$ is
well approximated by $\rcr \approx 0.82 \rho_{\mathrm{Pl}}$ up to
terms $O(\lp^2/a_{\rm min}^2)$, independently of the details of
the state and values of $p_\phi$. (For a universe with maximum
radius of a megaparsec, $\lp^2/a_{\rm min}^2 \approx 10^{-76}$.)
The density $\rmin$ at the recollapse point also agrees with the
value $(3/8 \pi G a^2_{\rm max})$ predicted by the classical
evolution to terms of the order $O(\lp^4/a_{\rm max}^4)$.
Furthermore the scale factor $a_{\rm max}$ at which recollapse
occurs in the quantum theory agrees to a very good precision with
the one predicted by the classical dynamics.

\b The trajectory obtained from effective Friedmann dynamics is in
excellent agreement with quantum dynamics \emph{throughout the
evolution.}  In particular, the maximum and minimum energy
densities predicted by the effective description agree with the
corresponding expectation values of the density operator $\hat
\rho \equiv \widehat{p_\phi^2/|p|^3}$ computed numerically.

\b The state remains sharply peaked for a \emph{very large number
of `cycles'.} Consider the example of a semi-classical state with
an almost equal relative dispersion in $p_\phi$ and $|v|_\phi$ and
peaked at a large classical universe of the size of a megaparsec.
When evolved, it remains sharply peaked with relative dispersion
in $|v|_\phi$ of the order of $10^{-6}$ \emph{even after $10^{50}$
cycles of contraction and expansion!} Any given quantum state
eventually ceases to be sharply peaked in $|v|_\phi$ (although it
continues to be sharply peaked in $p_\phi$). Nonetheless, the
quantum evolution continues to be deterministic and well-defined
for an infinite number of cycles. This is in sharp contrast with
the classical theory where the equations break down at
singularities and there is no deterministic evolution from one
cycle to the next. In this sense, in LQC  the $k$=1 universe is
{\it cyclic}, devoid of singularities. \emph{This non-singular
evolution holds for all states, not just the ones which are
semi-classical at late times.} There is no fine tuning of initial
conditions. Also, there is no violation of energy conditions.
Indeed, quantum corrections to the matter Hamiltonian do not play
any role in the resolution of the singularity. The standard
singularity theorems are evaded because the geometrical side of
the classical Einstein's equations is modified by the quantum
geometry corrections of LQC.

To summarize, the issues raised in section \ref{s1} have all been
answered in the FRW models. The main departures from the \WDW theory
occur due to quantum geometry effects of LQG. These effects are
small but dominate the Planck scale physics by creating an effective
repulsive force which can overwhelm gravitational attraction. While
these effects are small outside the Planck regime, in principle,
they could have accumulated and led to departures from general
relativity even in weak field regime on the very long time scales
that are relevant to cosmology. \emph{This does not happen.} While
they dominate the Planck regime, the quantum geometry effects die
extremely quickly outside this regime so that in the weak field
regime LQC is in excellent agreement with general relativity even on
the very large cosmological time scales.

\section{Discussion}
\label{s4}

In the last two section, I focused on a simple model to illustrate
how one might hope to solve the long standing problems of quantum
gravity using LQG. As I mentioned in the beginning of section
\ref{s2}, several other cosmological models have been analyzed.
Roughly, work to date can be divided into three categories.
\begin{itemize}
\item In the isotropic models with free scalar fields, with and
without cosmological constant, there are  detailed analytical as
well as numerical frameworks to answer all questions of physical
interest. In these models, there is only one non-trivial curvature
invariant ---the scalar curvature $R$-- and in the classical
theory it is related very simply to energy density. Therefore the
onset of the quantum epoch which signals departures from classical
general relativity can be described by critical values of
either the scalar curvature or the density \cite{aps3,apsv}.%
\footnote{Quantum geometry effects have also been shown to provide
a deterministic evolution in certain cyclic models \cite{svv}.}

\item For the Kasner model with anisotropies as well as models with
physically interesting potentials for scalar fields, the physical
sector of the theory can be constructed along the same lines.
However, the analysis of effective equations is still somewhat
incomplete and numerical analysis is still in infancy. In the
anisotropic models the key features of the isotropic models
---resolution of singularities and emergence of semi-classical
pre-big-bang branches--- persist. But new phenomena also emerge.
With massless scalar fields as sources, the scalar curvature $R$
is again determined by the matter density but there are also other
curvature invariants in particular because the Weyl tensor is no
longer zero. Now the quantum epoch is reached when any one of
these invariants reaches the Planck regime, whence multiple
bounces can occur \cite{kasner}. Density is no longer the
governing factor; rather it is space-time curvature. In models
with potentials, the scalar field does not always serve as
internal time globally. However, it is still possible to construct
the physical Hilbert space using group averaging \cite{dm} and
introduce relational observables. A global emergent time aids our
intuition immensely but is not essential in the construction of
the physical sector of the theory.

\item  In more complicated models ---black hole interiors
\cite{bh} and the so called midi-superspaces \cite{midi} which are
symmetry reduced but have an \emph{infinite} number of degrees of
freedom--- the Hamiltonian constraint has been written down. It
does not break down near the putative singularities signalling
that they are resolved. However, the physical sector of the theory
is yet to be constructed and numerical analysis has not yet been
undertaken. But there do not appear to be any unsurmountable
difficulties for these investigations to reach maturity in the
near future.

\end{itemize}

These advances are encouraging because they deal with the long
standing questions I discussed in section \ref{s1}. Furthermore,
in contrast to string theory, space-like singularities that are
resolved are of direct physical interest. However, the major
downside is that these advances are based on symmetry reduction
and the precise relation between these models and the full theory
is still to be spelled out. While significant efforts are being
made on this key problem \cite{bt}, I think we are still at a
preliminary stage largely because we do not have a clear candidate
for full LQG.

What lessons do these cosmological investigations have for full
LQG?

By now, the kinematic structure of full LQG is well controlled.
Key open issues involve dynamical issues. Work in this area has
been primarily focused on mathematical problems in full
generality. In the leading approaches, one first solves the
diffeomorphism constraint and then imposes the Hamiltonian
constraint on the resulting diffeomorphism invariant states. The
non-trivial achievement is that there are well-defined candidates
for constraint operators \cite{tt,ttbook}. Furthermore advances
related to Thiemann and Dittrich's `Master Constraint' program
\cite{master} strongly indicate that, for each admissible choice
of the master constraint, one would be able to construct the
physical sector of the theory. However, there is still a great
deal of freedom in the definition of the constraint operators and,
more importantly, the issue of existence of a sufficiently rich
semi-classical sector has remained largely open. More recently,
via their `algebraic quantum gravity program,'  Giesel and
Thiemann \cite{aqg} have introduced new strategies to address both
these issues. Here, the diffeomorphism and the Hamiltonian
constraints can be treated on an equal footing and imposed
simultaneously. This enables one to address the long standing
problem of recovering the classical constraint algebra and it is
in fact possible to recover it using suitably defined
semi-classical states \cite{toherent}. Thus there is ongoing
progress.

Nonetheless, fresh insights are needed to address key
\emph{physical} problems such as the fate of classical
singularities in full LQG and the detailed recovery of Einstein
dynamics in the classical limit. Since LQC has provided concrete
solutions to these problems in simple models, a useful strategy
would be to work `from bottom up' to less and less symmetric
models. For example, in the symmetry reduced systems, one
inevitably carries out a gauge fixing which provides a good
control on individual space-time geometries rather than
equivalence classes of them under diffeomorphisms. Similarly, the
standard description of the low energy world involves specific
space-time metrics. While one can translate both these
descriptions in a manifestly diffeomorphism invariant language,
the result would be quite cumbersome. The procedures used in
models can complement the more general and more systematic
programs that are being pursued in full LQG.

The idea would be to work one's way up by incorporating, at each
step, lessons learned from the symmetry reduced models. These
models suggest that, to address physically important issues, it
may be essential to restrict oneself to interesting sectors of the
theory ---finite, non-linear neighborhoods of the FRW solutions,
or of Minkowski space-time, in the phase space of general
relativity--- and exploit the additional structure such
restrictions make available. Secondly, in LQC one could get truly
valuable insights by analyzing in detail states which are
semi-classical in a suitable sense. By contrast, the discussion of
\emph{dynamics} in the full theory generally focuses on
`elementary states' ---the spin networks. These are analogous to
the `n-photon states' of the Maxwell theory, while the
semi-classical states are analogous to the coherent states. Both
span the full Hilbert space in the Maxwell theory and are
convenient in different regimes. Genuinely new insights could be
gained in interesting sectors of LQG if one revisits the issue of
constructing and solving the Hamiltonian constraint from a new
perspective. As in mini and midi superspace analyses, one could
exploit an astute gauge-fixing of (a part of) the diffeomorphism
constraint, and the extra structure provided by a basis of
semi-classical states. Results of LQC appear to provide valuable
guidelines not only for constructing the physical sector of the
theory along these lines but also for answering some of the most
challenging physical questions.

\section*{Acknowledgments} I have benefited from valuable discussions
with many colleagues I would like to thank especially Martin
Bojowald, Alex Corichi, James Hartle, Gary Horowitz, Veronika
Hubney, Jerzy Lewandowski, Donald Marolf, Tomasz Pawlowski,
Parampreet Singh, Thomas Thiemann, Kevin Vandersloot and
participants in the first st\"uckleberg workshop. This work was
supported in part by the NSF grants PHY99-07949 and PHY04-56913,
the Alexander von Humboldt Foundation, the Kramers Chair program
of the University of Utrecht, and the Eberly research funds of
Penn State.


\end{document}